\documentclass[aps, prd, onecolumn, tightenlines, notitlepage, superscriptaddress, nofootinbib, preprintnumbers, floatfix,showkeys,11pt]{revtex4-2}

\usepackage[normalem]{ulem}
\usepackage{amstext}
\usepackage{amssymb}
\usepackage{amsmath}
\usepackage{graphicx}
\graphicspath{{figures/}}
\usepackage{url}
\usepackage{color}
\usepackage{ulem}
\usepackage[utf8]{inputenc}
\pdfoutput=1
\usepackage{textcomp}
\usepackage{makecell}\usepackage{booktabs,siunitx,array,threeparttable}
\usepackage{epsfig,amsfonts,mathrsfs,amsmath,amssymb,graphicx,color,slashed,multirow}

\usepackage[x11names]{xcolor}
\usepackage[colorlinks]{hyperref}

\usepackage{textgreek} % To add greek letters to the text
\usepackage{caption, subcaption} % Better figure customization
\usepackage{booktabs} % Cool tables

\usepackage{xspace}

\definecolor{vdrgreen}{rgb}{0.0, 0.7, 0.0}

\definecolor{lightapricot}{rgb}{0.99, 0.84, 0.69}
\definecolor{nicered}{rgb}{0.7,0.1,0.1}
\definecolor{nicegreen}{rgb}{0.1,0.5,0.1}
\definecolor{coral}{rgb}{1.0, 0.5, 0.31}
\definecolor{blue(ncs)}{rgb}{0.0, 0.53, 0.74}
\definecolor{darkspringgreen}{rgb}{0.09, 0.45, 0.27}
\definecolor{seagreen}{rgb}{0.18, 0.55, 0.34}
\definecolor{cadmiumgreen}{rgb}{0.0, 0.42, 0.24}
\definecolor{chromeyellow}{rgb}{1.0, 0.65, 0.0}
\definecolor{darkturquoise}{rgb}{0.0, 0.81, 0.82}
\definecolor{denim}{rgb}{0.08, 0.38, 0.74}
\definecolor{purple(x11)}{rgb}{0.63, 0.36, 0.94}
\definecolor{red(ncs)}{rgb}{0.77, 0.01, 0.2}
\definecolor{ruddypink}{rgb}{0.88, 0.56, 0.59}
\definecolor{slateblue}{rgb}{0.42, 0.35, 0.8}
\definecolor{airforceblue}{rgb}{0.36, 0.54, 0.66}
\definecolor{orange(colorwheel)}{rgb}{1.0, 0.5, 0.0}

\makeatletter
    \newcommand{\colorboxed}[3][white]{\fcolorbox{#2}{#1}{\m@th$\displaystyle#3$}}
\makeatother
\DeclareMathOperator{\sinc}{sinc}

\AtBeginDocument{\hypersetup{citecolor=seagreen,linkcolor=seagreen,urlcolor=seagreen}}
\usepackage{appendix}

\begin{document}

\title{{\LARGE Confronting solutions of the Gallium Anomaly with reactor rate data}}

\author{Carlo Giunti}
\email{carlo.giunti@to.infn.it}
\affiliation{Istituto Nazionale di Fisica Nucleare (INFN), Sezione di Torino, Via P. Giuria 1, I--10125 Torino, Italy}

\author{Christoph A. Ternes}
\email{christoph.ternes@lngs.infn.it}
\affiliation{Istituto Nazionale di Fisica Nucleare (INFN), Laboratori Nazionali del Gran Sasso, 67100 Assergi, L’Aquila (AQ), Italy
}

\begin{abstract}
Recently, several models have been suggested to reduce the tension between Gallium and reactor antineutrino spectral ratio data
which is found in the framework of 3+1 active-sterile neutrino mixing.
Among these models, we consider the extensions of 3+1 mixing with a finite wavepacket size, or the decay of the heaviest neutrino $\nu_4$, or the possibility to have a broad $\nu_4$ mass distribution.
We consider the reactor antineutrino rate data and we show that these models cannot eliminate the tension between Gallium and
reactor rate data that is found in the 3+1 neutrino mixing framework.
Indeed, we show that the parameter goodness of fit remains small.
We consider also a model which explains the Gallium Anomaly with non-standard decoherence in the framework of three-neutrino mixing.
We find that it is compatible with the reactor rate data.
\end{abstract}
\maketitle

\section{Introduction}

The oscillations observed in
solar, atmospheric and long-baseline neutrino oscillation experiments
have established
the standard three-neutrino mixing framework
in which the three flavor neutrinos
$\nu_{e}$,
$\nu_{\mu}$,
$\nu_{\tau}$
are unitary superpositions of three light massive neutrinos
$\nu_{1}$,
$\nu_{2}$,
$\nu_{3}$
with respective masses
$m_{1}$,
$m_{2}$,
$m_{3}$.
(see, e.g., Ref.~\cite{ParticleDataGroup:2022pth}
and the recent global analyses in
Refs.~\cite{deSalas:2020pgw,Esteban:2020cvm,Capozzi:2021fjo}).
The oscillations are generated by
the two independent squared-mass differences
$\Delta m^2_{21} \approx 7.5 \times 10^{-5} \, \text{eV}^2$
and
$|\Delta m^2_{31}| \approx 2.5 \times 10^{-3} \, \text{eV}^2$
(with $\Delta m^2_{kj} \equiv m_{k}^2 - m_{j}^2$).
However,
anomalies observed in short-baseline (SBL) neutrino oscillation experiments
may require the existence of a larger squared-mass difference
$\Delta m^2_{\text{SBL}} \gtrsim 1 \, \text{eV}^2$,
which implies the extension of the standard three-neutrino mixing framework
to a model with more than three light massive neutrinos
(see, e.g., the recent reviews in
Refs.~\cite{Acero:2022wqg,Elliott:2023cvh,Zhang:2023zif}).
The simplest extension is the 3+1 model with a new massive neutrino
$\nu_{4}$
with mass $m_{4} \gtrsim 1 \, \text{eV}$,
such that
$\Delta m^2_{41} = \Delta m^2_{\text{SBL}} \gtrsim 1 \, \text{eV}^2$.
Since from the LEP measurements of the decay of the $Z$-boson~\cite{ALEPH:2005ab}
we know that there are only three active neutrinos,
in the flavor basis
the new neutrino is a sterile neutrino $\nu_{s}$,
which does not take part in weak interactions.
The sterile neutrino must be mostly mixed with the new massive neutrino,
in order to have a small perturbation of the three-neutrino mixing framework
which can explain the short-baseline anomalies
without spoiling the fit of
solar, atmospheric and long-baseline neutrino oscillation data:
\begin{equation}
\nu_{\alpha}
=
\sum_{k=1}^{4}
U_{\alpha k}
\nu_{k}
,
\label{eq:mixing}
\end{equation}
for $\alpha=e,\mu,\tau,s$,
where $U$ is the unitary $4\times4$ mixing matrix
such that
$|U_{e4}|^2+|U_{\mu4}|^2+|U_{\tau4}|^2=1-|U_{s4}|^2\approx1$.

Most puzzling is the short-baseline Gallium Anomaly (GA),
which is a deficit of events observed in Gallium source experiments
(GALLEX~\cite{GALLEX:1997lja,Kaether:2010ag},
SAGE~\cite{Abdurashitov:2005tb,SAGE:2009eeu}, and
BEST~\cite{Barinov:2021asz,Barinov:2022wfh})
with respect to the rate expected in the three-neutrino mixing framework.
Since the Gallium Anomaly deficit is relatively large,
it is in tension with the measurements of short-baseline reactor neutrino experiments
in the framework of 3+1 neutrino mixing~\cite{Giunti:2022btk}.

It has been proposed to relieve this tension by introducing new effects
which damp the oscillations in short-baseline reactor neutrino experiments:
a quantum mechanical wavepacket effect~\cite{Arguelles:2022bvt,Hardin:2022muu},
the decay of the new mass state $\nu_{4}$~\cite{Hardin:2022muu},
and a broad mass distribution for $\nu_{4}$~\cite{Banks:2023qgd}. 
The oscillation damping reduces the bounds obtained from the ratios of events
measured at different distances in the short-baseline reactor experiments
NEOS~\cite{NEOS:2016wee,RENO:2020hva},
DANSS~\cite{DANSS:2018fnn,DANSS-ICHEP2022},
PROSPECT~\cite{PROSPECT:2018dtt,PROSPECT:2020sxr}, and
STEREO~\cite{STEREO:2018rfh,STEREO:2019ztb},
relieving the tension between the results of these experiments
and the Gallium Anomaly~\cite{Hardin:2022muu,Banks:2023qgd}.

In this paper we show that, however,
the new damping effects cannot relieve the tension between the
results of short-baseline reactor neutrino rate experiments
and the Gallium Anomaly.
The rate experiments, summarized in Table~4 of Ref.~\cite{Giunti:2021kab},
measured the total short-baseline reactor neutrino event rates.
In 2011 the comparison with the event rates expected from the theoretical calculation of the reactor electron antineutrino flux
(the HM model of
Huber~\cite{Huber:2011wv}
and
Mueller \textit{et al}~\cite{Mueller:2011nm})
generated the Reactor Antineutrino Anomaly (RAA)~\cite{Mention:2011rk},
which is a deficit of events with respect to the prediction.
However,
this deficit is smaller than that of the Gallium Anomaly
and it is in tension with it~\cite{Giunti:2022btk}.
Moreover,
new reactor electron antineutrino flux calculations
(the EF model of
Estienne, Fallot, \textit{et al}~\cite{Estienne:2019ujo} and its revision~\cite{Perisse:2023efm}
and
the KI model of
Kopeikin \textit{et al.}~\cite{Kopeikin:2021ugh})
decreased the Reactor Antineutrino Anomaly~\cite{Berryman:2020agd,Giunti:2021kab}
and increased the tension with the Gallium Anomaly~\cite{Giunti:2022btk}.

The new damping effects do not reduce significantly the tension between
the reactor rates bound and the Gallium Anomaly,
because the neutrino oscillations relevant for the reactor rates
are already almost completely averaged in the 3+1 model for the values of
$\Delta m^2_{41} \gtrsim 1 \, \text{eV}^2$
which fit the Gallium Anomaly.

We consider also the explanation of the Gallium Anomaly
proposed in Ref.~\cite{Farzan:2023fqa}
through non-standard decoherence effects in the framework of three-neutrino mixing
and we obtain the condition for its compatibility with the reactor rate data.

The plan of the paper is to discuss
the wavepacket effect in Section~\ref{sec:wavepacket},
$\nu_{4}$ decay in Section~\ref{sec:decay},
a broad $\nu_{4}$ mass distribution in Section~\ref{sec:broad}, and
the three-neutrino scenario with non-standard decoherence effects
in Section~\ref{sec:decoherence}.
Finally, in Section~\ref{sec:conclusions}
we present a summary and conclusions.

\section{The wavepacket effect}
\label{sec:wavepacket}

\begin{figure}
    \includegraphics[width=0.45\textwidth]{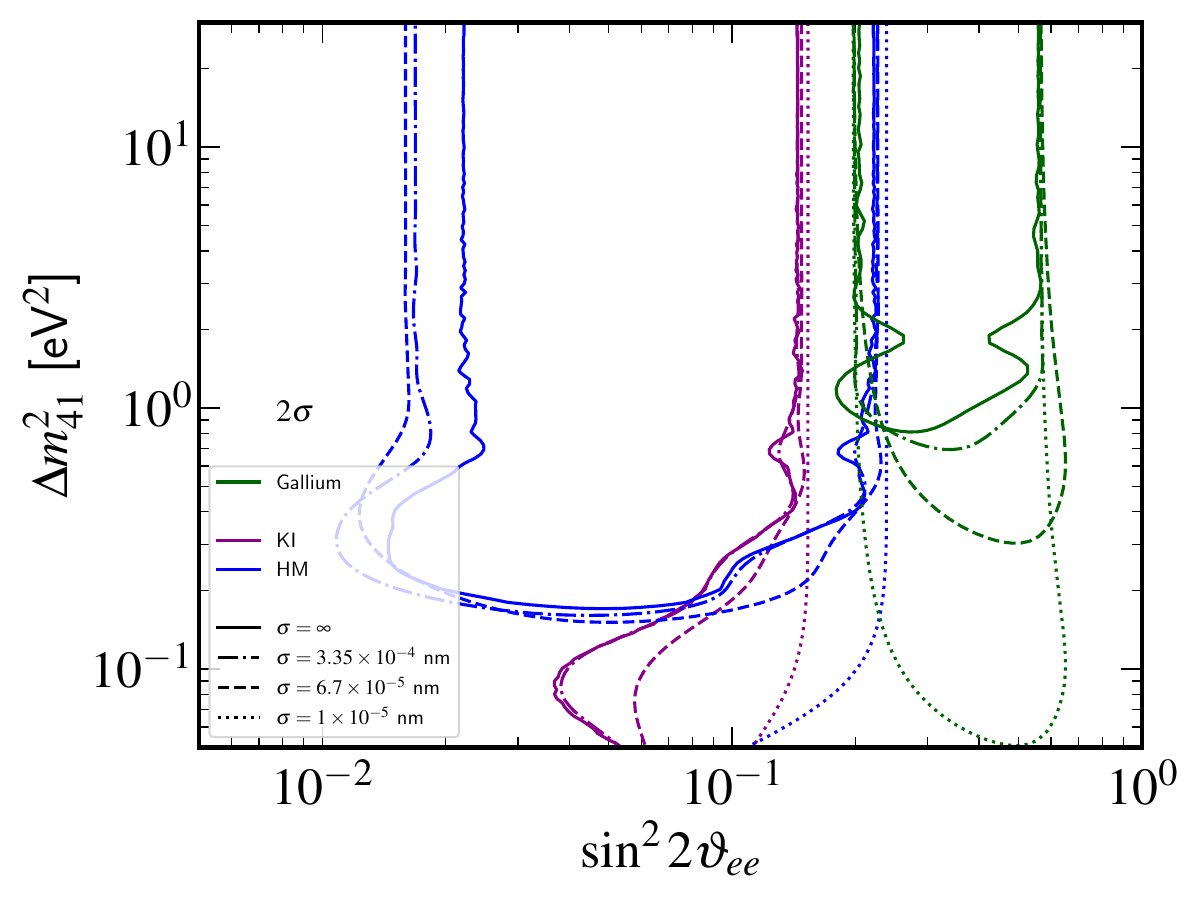}
\includegraphics[width=0.45\textwidth]{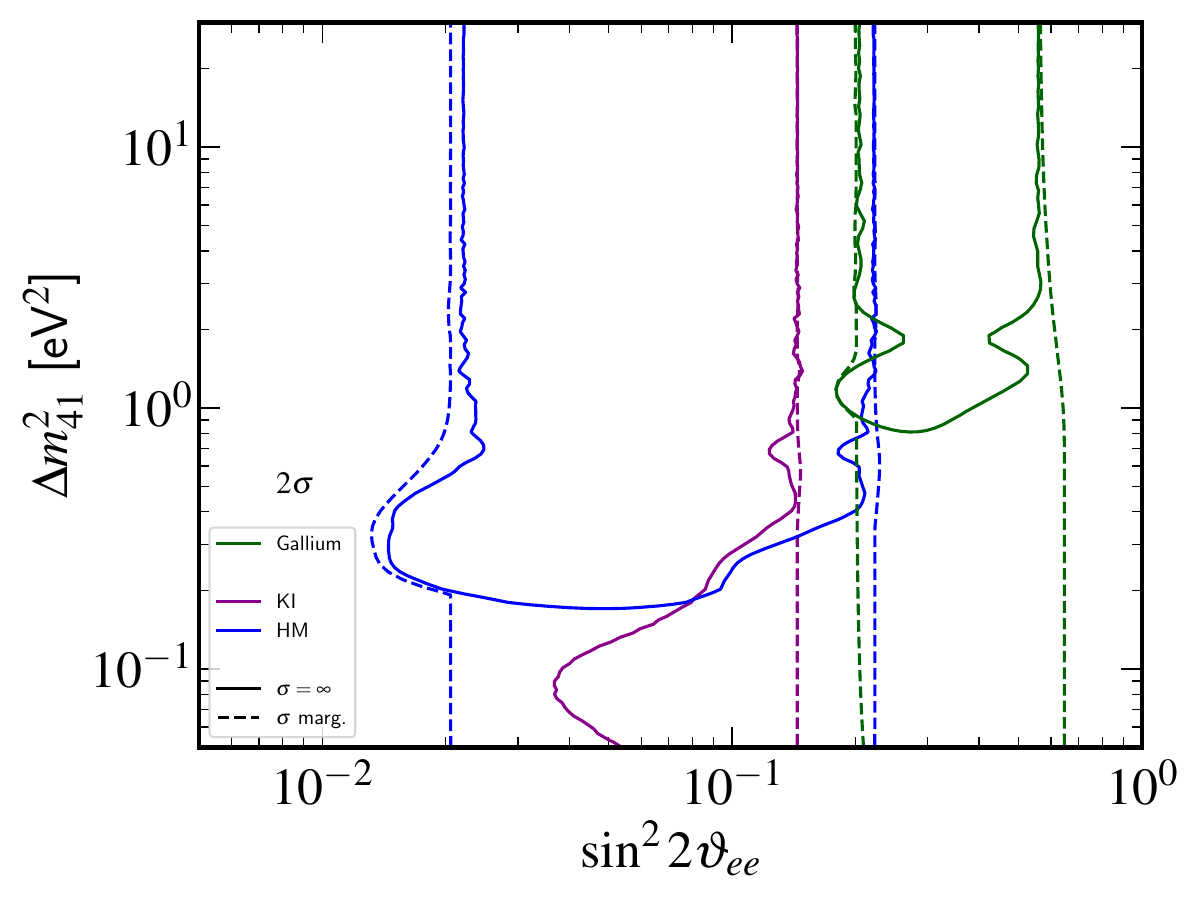}  %caption
    %label
\caption{In the left panel we show the allowed regions of parameter space at 2$\sigma$ from the analyses of reactor rate data using the KI (magenta) and HM (blue) fluxes and of the Gallium data (green) for some selected values for the wavepacket width $\sigma$. The contours for each case are plotted with respect to the local minimum obtained with the wavepacket widths indicated in the legend. In the right panel we show the allowed regions after marginalizing over $\sigma$.}
\label{fig:wavepacket}
\end{figure}

In this Section we discuss the effects of a small wavepacket width. This scenario was considered in Ref.~\cite{Arguelles:2022bvt} analyzing spectral shape data from reactor experiments and the Gallium data. In Ref.~\cite{Hardin:2022muu} it was considered withtin a global fit to neutrino oscillation data. The effective neutrino oscillation probability at very short baselines is given by
\begin{equation}
    P_{ee}^{\text{WP}} = 1 - \frac{1}{2}\sin^22\vartheta_{ee}
    \left[ 1 - 
    \cos\left(\frac{\Delta m_{41}^2L}{2E}\right) e^{-\left(\frac{L}{L_{\text{coh}}}\right)^2}
    \right]\,,
    \label{eq:osc_WP}
\end{equation}
where
$ \sin^2 2\vartheta_{ee} = 4 |U_{e4}|^2 ( 1 - |U_{e4}|^2 ) $,
$E$ is the neutrino energy,
and $L$ is the source-detector distance.
The coherence length given by
\begin{equation}
    L_{\text{coh}} = \frac{4 \sqrt{2} E^2}{|\Delta m_{41}^2|}\sigma\,,
\end{equation}
where $\sigma$ is the width of the wavepacket in coordinate space. The size of the wavepacket has been recently estimated in Ref.~\cite{Akhmedov:2022bjs}. The result of this analysis shows that the wavepacket is very large
($\sigma \simeq 200 \, \text{nm}$ for reactor neutrinos and
$\sigma \simeq 1400 \, \text{nm}$ for Gallium experiments)
and does not have any effect on neutrino oscillations in reactor and Gallium experiments. Even if this calculation is not taken into account, the required wavepacket size of Ref.~\cite{Hardin:2022muu}
($\sigma \approx 6.7 \times 10^{-5} \, \text{nm}$)
is in tension with the phenomenological bounds obtained in Refs.~\cite{deGouvea:2020hfl,deGouvea:2021uvg}: $\sigma > 2.1 \times 10^{-4} \, \text{nm}$ at 90\% confidence level (C.L.).
However, in the current analysis of this paper we do not take into account any prior information on the wavepacket size and we consider $\sigma$ as an unbounded parameter.

The results of our analyses of reactor and Gallium data for neutrino oscillations with finite wavepacket size are shown in Fig.~\ref{fig:wavepacket}. 
For simplicity we consider only the HM and KI reactor antineutrino flux models.
The HM model is the original model of the Reactor Antineutrino Anomaly and the KI model is its revision taking into account the new measurements in Ref.~\cite{Kopeikin:2021ugh}.
In the case of Gallium data the choice of cross section model can have an impact on the tension with reactor rate data~\cite{Giunti:2022btk}. In this paper we consider for simplicity only the Bahcall cross section model~\cite{Bahcall:1997eg}.

In the left panel of Fig.~\ref{fig:wavepacket} we plot the allowed regions for some selected values of $\sigma$: the best-fit value from Ref.~\cite{deGouvea:2021uvg}, namely $\sigma = 3.35\times10^{-4}$~nm (dashed-dotted lines), the best-fit value from Ref.~\cite{Hardin:2022muu}, namely $\sigma = 6.7\times10^{-5}$~nm (dashed lines), and the even smaller value $\sigma = 1\times10^{-5}$~nm (dotted lines). 
Note that the best-fit wavepacket size of Ref.~\cite{deGouvea:2021uvg} has little effect on the allowed regions when compared to those in the standard 3+1 analysis. When allowing for smaller wavepackets, even smaller than the best-fit value from Ref.~\cite{Hardin:2022muu}, the upper bound on $\sin^2 2\vartheta_{ee}$ of the reactor allowed region,
which is attained at large values of $\Delta m_{41}^2$,
is almost unaffected.
For very small wavepacket sizes,
the Gallium allowed region is extended to low values of $\Delta m_{41}^2$,
but it still requires values of $\sin^2 2\vartheta_{ee}$ larger than the reactor upper bound.
Therefore, the wavepacket effect cannot relieve the tension between the reactor rate data and the Gallium data.

\begin{table}[t]
\centering
\begin{tabular}{|c|c|c|c|c|}
\hline
  & 3+1 & ~~~wavepacket~~~ & ~~~decay~~~ & ~~~broad $\nu_4$~~~\\
\hline
$~~~\textrm{GoF}_{\textrm{PG}}$(GA+RR(HM))~~~ & $~~~2.1\times10^{-2}~~~$& $~~~2.3\times10^{-2}~~~$& $~~~2.1\times10^{-2}~~~$& $~~~2.1\times10^{-2}~~~$\\
\hline
$~~~\textrm{GoF}_{\textrm{PG}}$(GA+RR(KI))~~~ & $1.1\times10^{-3}$& $2.2\times10^{-3}$& $2.2\times10^{-3}$& $2.2\times10^{-3}$\\
\hline
\end{tabular}
\caption{The parameter goodness of fit  obtained for the models under consideration for the combination of Gallium (GA) data with reactor rate (RR) data using the HM flux (first row) and the KI flux (second row).}
\label{tab:gof}
\end{table}

In the right panel of Fig.~\ref{fig:wavepacket} we show the region obtained after marginalizing over $\sigma$. One can see that the wavepacket effect mainly affects the regions of parameter space at low values of $\Delta m_{41}^2$. It is clear from the figure that the tension between the reactor and Gallium allowed regions persists. 
In order to quantify the tension we compute the parameter goodness of fit~\cite{Maltoni:2003cu} for the analysis including the wavepacket effect and compare it to that obtained with the 3+1 analysis. 
The results are shown in Tab.~\ref{tab:gof},
where one can see that the inclusion of the wavepacket effect has very little impact on the parameter goodness of fit in the analysis with the HM flux model.
Instead, there is an improvement in the analysis with the KI flux model, but the tension remains much worse than for the HM flux model.
Therefore, we conclude that
the wavepacket effect cannot eliminate
the tension between the reactor rate data and the Gallium data.

Note that, in principle, the wavepacket of neutrinos in Gallium experiments does not need to have the same size as that in reactor neutrino experiments.
We chose the same value for both data sets for illustration, 
but one can confront the reactor and Gallium allowed regions
with different values of $\sigma$ in Fig.~\ref{fig:wavepacket}
and see that they are anyway in tension.

\section{$\nu_{4}$ decay}
\label{sec:decay}

\begin{figure}
    \includegraphics[width=0.45\textwidth]{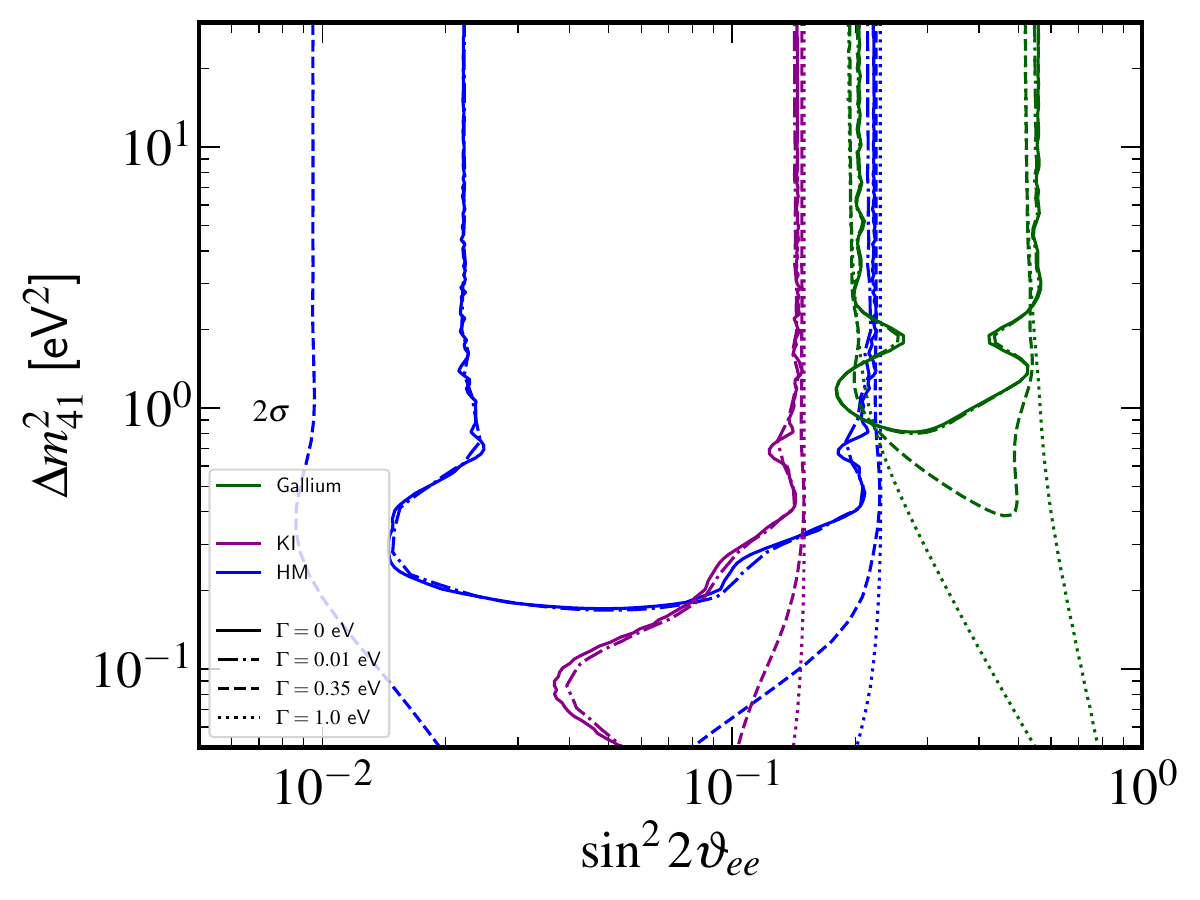}
    \includegraphics[width=0.45\textwidth]{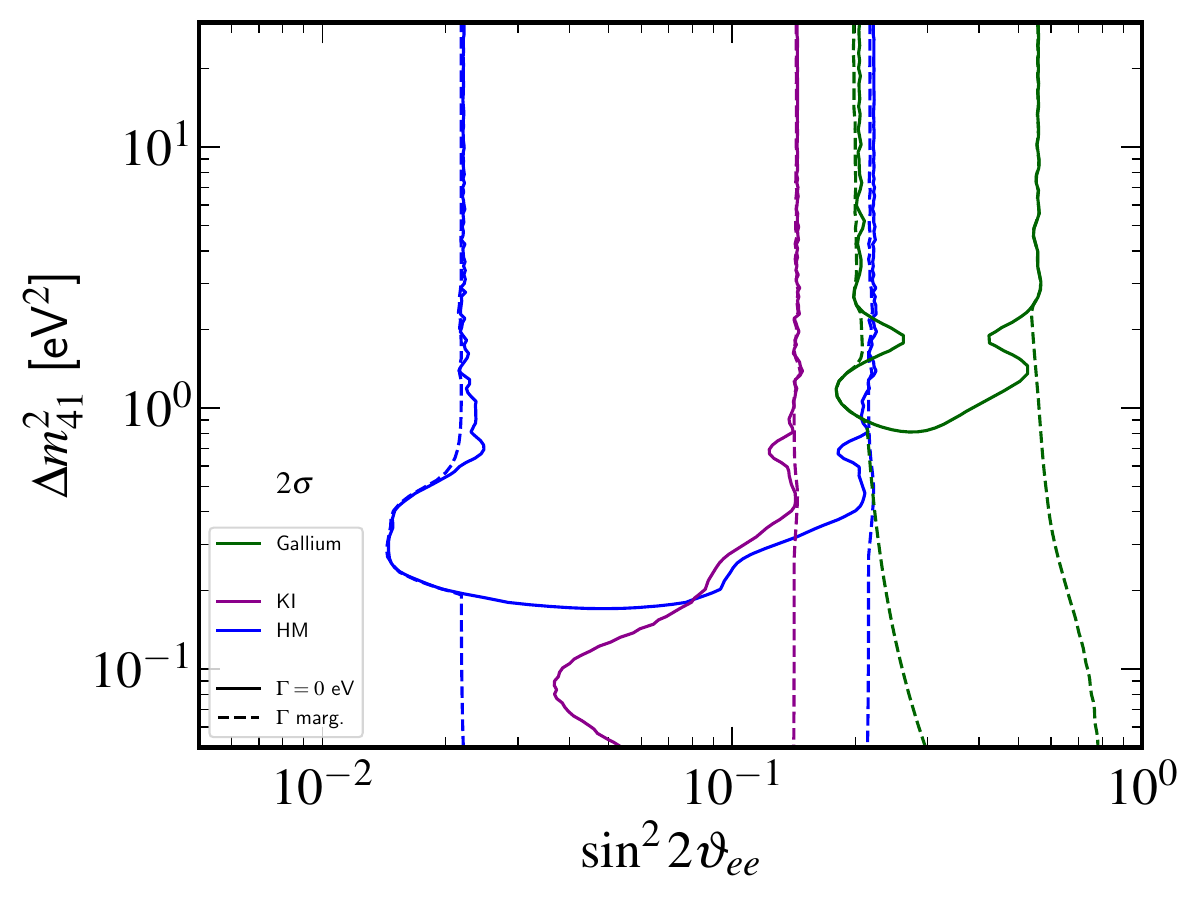}
    %caption
    %label
\caption{In the left panel we show the allowed regions of parameter space at 2$\sigma$ from the analyses of reactor rate data using the KI (magenta) and HM (blue) fluxes and of the Gallium data (green) for some selected values of the decay width $\Gamma$ of $\nu_4$. The contours are plotted with respect to the minimum obtained using the decay widths as indicated in the legend. In the right panel we show the allowed regions after marginalizing over $\Gamma$.}
\label{fig:decay}
\end{figure}

We turn our attention now to the second scenario considered in Ref.~\cite{Hardin:2022muu}, the possible decay of $\nu_4$. As in the wavepacket scenario this model improves the fit of neutrino oscillation data according to Ref.~\cite{Hardin:2022muu}. In this case the neutrino oscillation probability at short baselines is given by 
\begin{equation}
    P_{ee}^{\textrm{dec.}} = 2|U_{e4}|^2(1-|U_{e4}|^2)e^{-\frac{\Gamma m_4L}{2E}}\cos\left(\frac{\Delta m_{41}^2L}{2E}\right) + |U_{e4}|^4e^{-\frac{\Gamma m_4L}{E}} + (1-|U_{e4}|^2)^2\,,
    \label{eq:osc_dec}
\end{equation}
where $\Gamma = 1/\tau$ is the decay width and $\tau$ is the lifetime of $\nu_4$. Since $\nu_4$ has a much larger mass than the other neutrinos we approximate $m_4 \simeq \sqrt{\Delta m_{41}^2}$. 

We fit Eq.~\ref{eq:osc_dec} assuming that $\Gamma$ is a completely free parameter, not taking into account any possible bounds. The results are shown in Fig.~\ref{fig:decay}. In the left panel we show the results for some selected values of $\Gamma$, while in the right panel we show the allowed regions after marginalization over the decay width. As can be seen when fixing $\Gamma = 0.35$~eV (which is the best-fit value from Ref.~\cite{Hardin:2022muu}), the allowed region of the reactor rate analysis with the HM fluxes opens up towards lower masses, while in the case of the KI flux the bound at small masses becomes less stringent.
However, the upper limit on $\sin^22\vartheta_{ee}$,
which is reached for large values of $\Delta m_{41}^2$,
is unaffected by the decay. This remains true when allowing for larger values of $\Gamma$ and also when marginalizing over it, as shown in the right panel of the figure.

The effect of the decay on the Gallium region is also not helping in reducing the tension with the reactor rate data. The allowed parameter space is larger than in the standard 3+1 case, but the new region requires even larger mixing angles.
Therefore, the tension between the reactor rate data and the Gallium data is not eliminated by the decay.  
This can be seen from the values of the parameter goodness of fit in Tab.~\ref{tab:gof},
where one can see that considering the HM flux model
the inclusion of the decay does not change the parameter goodness of fit
with respect to that obtained with the 3+1 analysis.
There is instead an improvement in the analysis with the KI flux model,
but the tension remains much worse than for the HM flux model.
Hence, we conclude that
the addition of $\nu_4$ decay to the 3+1 model cannot eliminate
the tension between the reactor rate data and the Gallium data.

\section{Broad $\nu_{4}$ mass distribution}
\label{sec:broad}

\begin{figure}
    \includegraphics[width=0.45\textwidth]{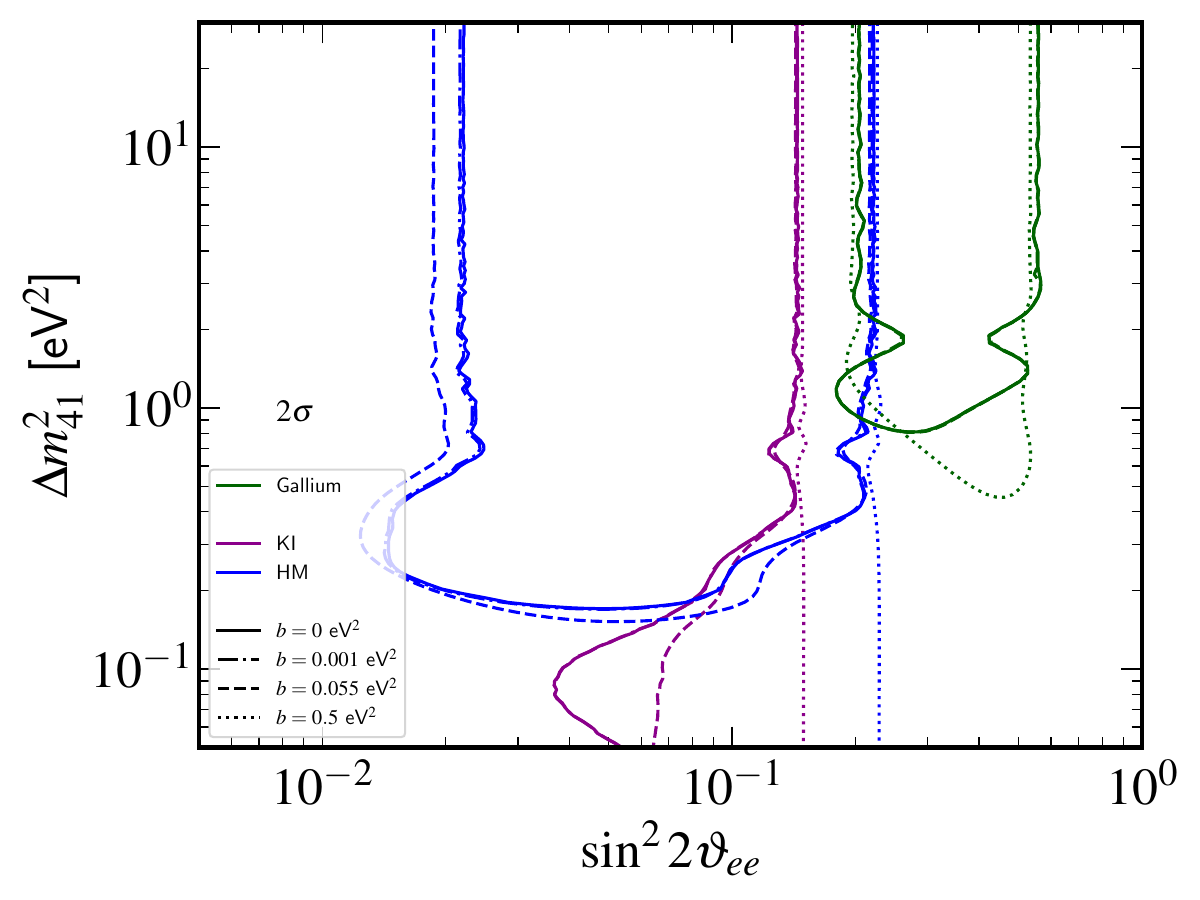}
    \includegraphics[width=0.45\textwidth]{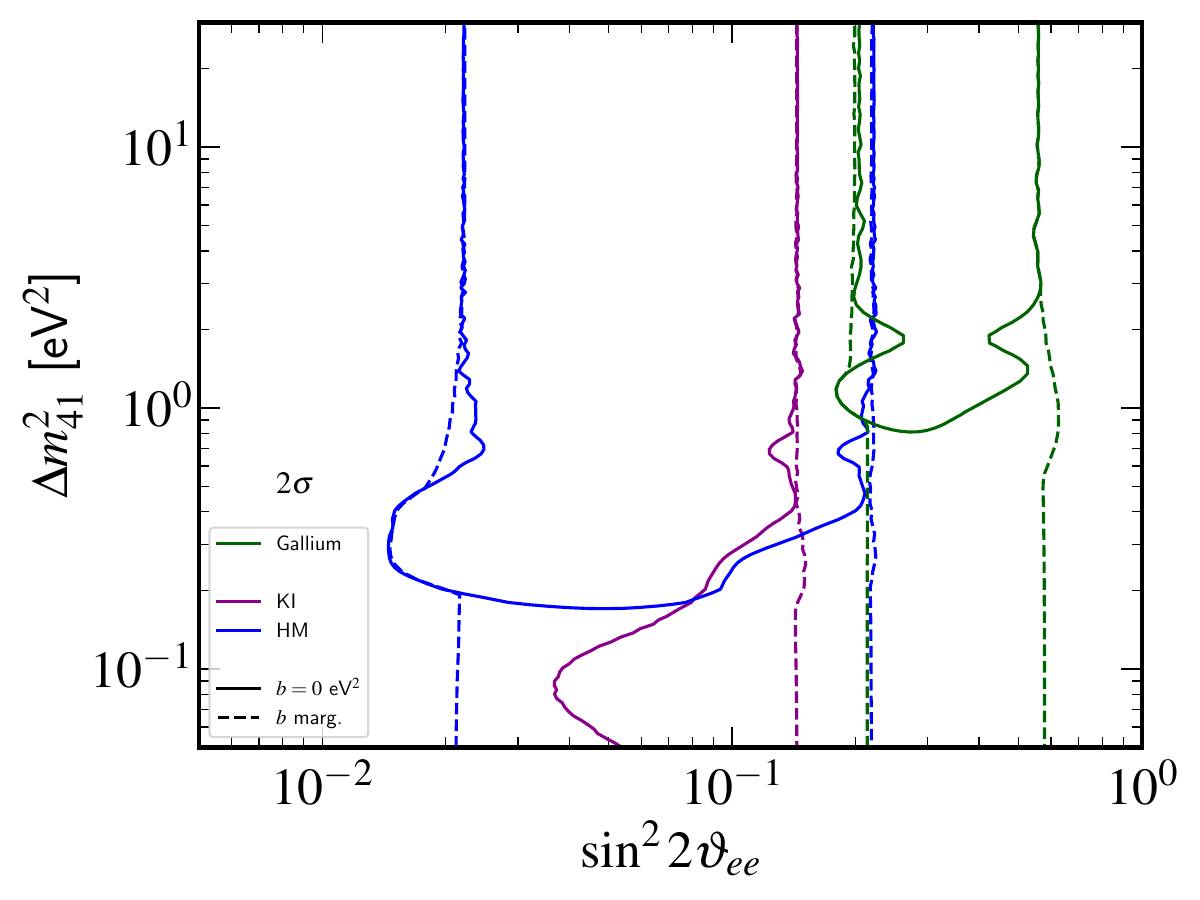}
    %caption
    %label
\caption{In the left panel we show the allowed regions of parameter space at 2$\sigma$ from the analyses of reactor rate data using the KI (magenta) and HM (blue) fluxes and of the Gallium data (green) for some selected values of the breadth $b$ of a broad $\nu_4$ mass distribution. The contours are plotted with respect to the minimum obtained using the breadth values indicated in the legend. In the right panel we show the allowed regions after marginalizing over $b$.}
\label{fig:broad}
\end{figure}

In this Section we consider a model with a broad $\nu_4$ mass distribution, which was developed in Ref.~\cite{Banks:2022gwq} and applied to Gallium and reactor spectral ratio data in Ref.~\cite{Banks:2023qgd}. In this model it is assumed that the fourth mass-eigenstate can be modeled as a state with a central mass squared $m_4^2$ and finite breadth $b$. In this scenario the effective neutrino oscillation probability at short baselines is given by 
\begin{equation}
    P_{ee}^{b} = \left(1 + \left(\sinc\left(\frac{b L}{4E}\right)-1\right)|U_{e4}|^2\right)^2 -4|U_{e4}|^2 \left(1 - |U_{e4}|^2\right)\sin^2{\left(\frac{\Delta m_{41}^2 L}{4E}\right)} \sinc\left(\frac{b L}{4E}\right)\,,
    \label{eq:osc_broad}
\end{equation}
where $\sinc(x)=\frac{\sin(x)}{x}$. Note that $\sinc(0) = 1$ and therefore for $b=0$ the standard 3+1 neutrino oscillation probability is recovered. 

The results of the fit of Eq.~\eqref{eq:osc_broad} are shown in Fig.~\ref{fig:broad}. As in the previous cases, we show the contours for selected values of $b$ in the left panel and the contours obtained after marginalizing over $b$ in the right panel.
As can be seen from the left panel of Fig.~\ref{fig:broad},
the best-fit value of $b$ from Ref.~\cite{Banks:2023qgd} ($b=0.055\,\text{eV}^2$)
has little effect on the allowed regions.
For the much larger value $b=0.5\,\text{eV}^2$
the reactor rate data give upper bounds for $\sin^22\vartheta_{ee}$
which are independent of $\Delta m_{41}^2$
and have the same values as those of the 3+1 analysis for large values of $\Delta m_{41}^2$
in both analyses with the HM and KI flux models.
The Gallium region is slightly extended to lower values of $\Delta m_{41}^2$
without changing significantly the lower bound for $\sin^22\vartheta_{ee}$,
which remains in tension with the reactor upper bound.

The right panel in Fig.~\ref{fig:broad}
shows that also after marginalizing over $b$ the tension between the reactor and Gallium allowed regions persists\footnote{Note that the differences of the marginalized Gallium allowed region with that in Fig.~3 of Ref.~\cite{Banks:2023qgd} is due to the fact that we marginalize over $b$ freely, while in Ref.~\cite{Banks:2023qgd} the authors require $b<2m_4^2$.}.
This is confirmed by the values of the parameter goodness of fit in Tab.~\ref{tab:gof},
which are the same as those obtained in the $\nu_4$ decay analysis.
Therefore,
we conclude that also a broad $\nu_{4}$ mass distribution
cannot eliminate
the tension between the reactor rate data and the Gallium data.

\section{Non-standard decoherence effects}
\label{sec:decoherence}

\begin{figure}
    \includegraphics[width=0.45\textwidth]{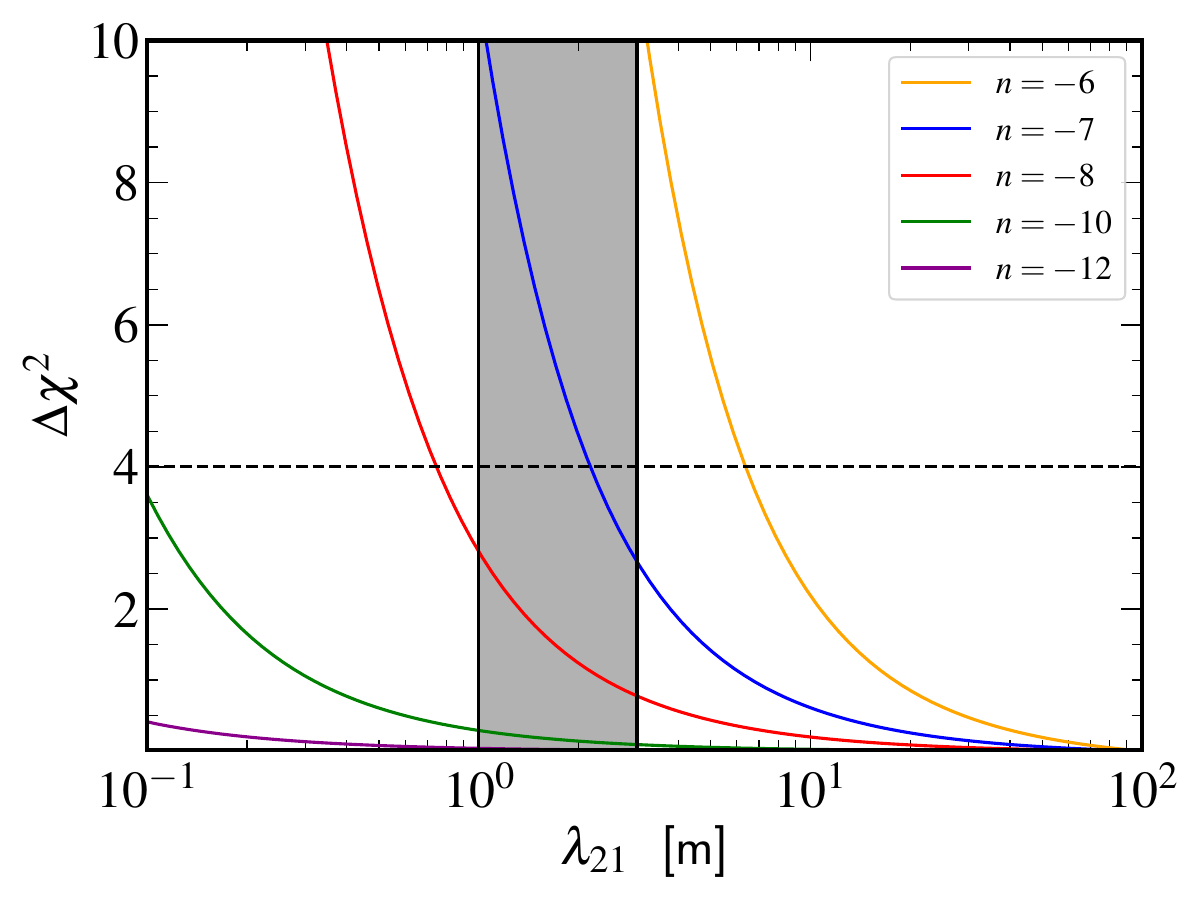}
    %caption
    %label
\caption{Constraints on the decoherence length $\lambda_{21}$ from reactor rate data using the KI flux model. The shaded region is the approximate preferred region of Ref.~\cite{Farzan:2023fqa} for the explanation of the Gallium Anomaly.  The horizontal dashed line indicates the $2\sigma$ confidence level.}
\label{fig:decoherence}
\end{figure}

An alternative solution, which does not require a fourth neutrino, has been proposed in Ref.~\cite{Farzan:2023fqa}. In this reference the authors explain the Gallium Anomaly
in the framework of three neutrino mixing with a non-standard loss of coherence of the low-energy
neutrinos in Gallium source experiments. The neutrino oscillation probability in this scenario is given by 
\begin{equation}
    P_{ee}^{\textrm{deco.}} = 1 - \frac{1}{2}\sin^22\theta_{13} - \frac{1}{2}\cos^4\theta_{13}\sin^22\theta_{12}\left(1-e^{-\gamma_{21}L}\right)\,,
    \label{eq:osc_decoherence}
\end{equation}
where
\begin{equation}
    \gamma_{21}=\frac{1}{\lambda_{21}}\left(\frac{E}{E_0}\right)^n
\end{equation}
is the decoherence parameter.
Following Ref.~\cite{Farzan:2023fqa}, we set $E_0 = 0.75$~MeV.
In Eq.~\eqref{eq:osc_decoherence} it is assumed that there is full decoherence in the 13-sector,
according to the fit of Gallium data in Ref.~\cite{Farzan:2023fqa}.
The authors of Ref.~\cite{Farzan:2023fqa} show that the Gallium Anomaly could be explained with energy dependencies $n\lesssim-2$ and argue that extreme dependencies $n\lesssim-10$
do not affect the observed oscillations of higher-energy solar and reactor neutrinos.
Such extreme energy dependencies have never been considered before~\cite{Lisi:2000zt,Fogli:2007tx,Coloma:2018idr,Guzzo:2014jbp,Carrasco:2018sca,Gomes:2020muc,Stuttard:2020qfv,DeRomeri:2023dht}.

We calculate bounds on the decoherence length $\lambda_{21}$ using reactor rate data.
For simplicity, this time we consider only the KI flux model
(the results with the HM model are similar).
The marginal $\Delta\chi^2=\chi^2-\chi^2_{\text{min}}$
as a function of $\lambda_{21}$
is shown in Fig.~\ref{fig:decoherence}
for several choices of $n$.
The shaded region in the figure is the approximate preferred region of Ref.~\cite{Farzan:2023fqa} for the explanation of the Gallium Anomaly
with $n\lesssim-2$.
One can see that at $2\sigma$ this explanation of the Gallium Anomaly is not in conflict with reactor rate data as long as $n\lesssim-7$.
More extreme energy dependencies, such as $n=-12$ considered in
Ref.~\cite{Farzan:2023fqa},
cannot be tested with reactor rate data.
Therefore,
we conclude that the explanation of the Gallium Anomaly proposed in Ref.~\cite{Farzan:2023fqa}
is allowed by the reactor rate data.

\section{Conclusions}
\label{sec:conclusions}

In this paper we have examined three models which
have been proposed~\cite{Arguelles:2022bvt,Hardin:2022muu,Banks:2023qgd} to relieve the tension between Gallium and reactor experiments
that is found in the 3+1 neutrino mixing framework~\cite{Giunti:2022btk}.
While this is true when considering reactor spectral ratio data~\cite{Arguelles:2022bvt,Hardin:2022muu,Banks:2023qgd},
we have shown that
these models are in strong tension with reactor rate data. 
In Tab.~\ref{tab:gof} we summarize the parameter goodness of fit~\cite{Maltoni:2003cu}
of the analyses of Gallium data and reactor rate data for each model and we compare it with the one obtained from the standard 3+1 analysis.
As can be seen, the goodness of fit obtained using the HM reactor antineutrino flux is practically the same in the three models as in the 3+1 analysis.
When considering the KI flux instead, the goodness of fit is better than in the standard 3+1 case. Unfortunately, it has still a very low value.
Therefore, we conclude that the three models that we have considered
cannot eliminate the tension between Gallium and reactor rate data
that is found in the 3+1 neutrino mixing framework.

We considered also, in Section~\ref{sec:decoherence},
a model with non-standard decoherence proposed in Ref.~\cite{Farzan:2023fqa}
to explain the Gallium Anomaly in the three-neutrino mixing framework.
We have shown that it is compatible with the reactor rate data
if the power $n$ of the energy dependence of the decoherence parameter
is $n\lesssim-7$.
Therefore, the explanation of the Gallium Anomaly with $n\lesssim-10$
proposed in Ref.~\cite{Farzan:2023fqa}
is allowed by the reactor rate data\footnote{We did not calculate the parameter goodness of fit for the decoherence model
because there is no tension for $n\lesssim-7$.}.

Reactor rate data are often discarded in analyses arguing that the results are model dependent. We disagree with this argument. While it is true that there is a model dependence, the differences among the results with different antineutrino flux models are not huge and they give upper bounds on the 3+1 oscillation parameter $\sin^22\vartheta_{ee}$~\cite{Giunti:2021kab} which must be taken into account.
A good indication that the reactor flux models are more reliable now than some years ago is that there is good agreement~\cite{Giunti:2021kab} between the predictions of the EF~\cite{Estienne:2019ujo} and KI~\cite{Kopeikin:2021ugh} flux models, which have been obtained using two completely different techniques.
Many results in particle physics are obtained under certain assumptions and are not criticized or discarded for being model-dependent. Neither should reactor rate data be ignored.

Regarding the Gallium Anomaly, several oscillation explanations are in tension with the reactor rate data~\cite{Forero:2022skg,Giunti:2022btk,Arguelles:2022bvt,Hardin:2022muu,Banks:2023qgd}. Other explanations, with Standard Model and beyond the Standard Model physics, have been suggested in Refs.~\cite{Giunti:2022xat,Brdar:2023cms}, but some of the Standard Model explanations are already excluded by the new measurements of $^{71}\text{Ge}$ decay presented in Ref.~\cite{Collar:2023yew}. As of now, the Gallium Anomaly remains a mystery.

%\bibliographystyle{utphys}
%\bibliography{bibliography,main}  

\providecommand{\href}[2]{#2}\begingroup\raggedright\endgroup

\end{document}